\documentclass[a4paper,11pt]{article}
\usepackage{pos}

\title{Extrapolating semileptonic form factors using 
Bayesian-inference fits regulated by unitarity and analyticity}
\author[a,b]{J.M.~Flynn}
\author[a,b,c]{A.~Jüttner}
\author[c]{J.T.~Tsang}

\affiliation[a]{School of Physics and Astronomy, University of Southampton, Southampton, SO17 1BJ, UK}
\affiliation[b]{STAG Research Centre, University of Southampton, Southampton, SO17 1BJ, UK}
\affiliation[c]{Theoretical Physics Department, CERN, Geneva, Switzerland}

\emailAdd{j.m.flynn@soton.ac.uk}
\emailAdd{Andreas.Juttner@cern.ch}
\emailAdd{j.t.tsang@cern.ch}

\abstract{We discuss our recently proposed~\cite{Flynn:2023qmi}
  model-independent framework for fitting hadronic form-factor data, which are
  often only available at discrete kinematical points, using parameterisations
  based on unitarity and analyticity. The accompanying dispersive bound on the
  form factors (unitarity constraint) is used to regulate the ill-posed fitting
  problem and allow model-independent predictions over the entire physical
  range. Kinematical constraints, for example for the vector and scalar form
  factors in semileptonic meson decays, can be imposed exactly. The core
  formulae are straight-forward to implement with standard math libraries. We
  demonstrate the method for the exclusive semileptonic decay $B_s\to K\ell\nu$,
  an example requiring one to use a generalisation of the original
  Boyd~Grinstein~Lebed (BGL) unitarity constraint. We further present a first
  application of the method to $B \to D^*\ell \nu$ decays.}

\FullConference{The 40th International Symposium on Lattice Field Theory (Lattice 2023)\\
July 31st - August 4th, 2023\\
Fermi National Accelerator Laboratory\\}

\begin{document}
\maketitle

\section{Introduction}
In the study of semileptonic $B_{(s)}$-meson decays on the lattice one still faces the difficulty of reconciling all relevant physical scales at the same time, while keeping systematic uncertainties at bay.\footnote{For a recent
  review of $B_{(s)}$-physics from lattice QCD we refer the reader
  to Ref.~\cite{Tsang:2023nay}.} For instance, the heavy mass of the $b$ quark and the requirement to induce large spatial final-state momenta both require fine lattice spacings. There is also the requirement of large physical lattice volumes to limit finite-size effects. 
Besides these regulator-dependent limitations, there is also the problem of the deteriorating signal-to-noise ratio when including data for larger final-state momenta, data which is required to cover 
a larger fraction of the kinematically allowed momentum transfer between the initial and final state hadrons. 

While accommodating all the above constraints remains a long-term goal, a widely
used strategy in the meantime is to create lattice data for (near-)physical
simulation parameters (in particular quark masses), and to restrict predictions
to relatively large momentum transfer $q^2$. In order to make predictions for
hadronic form factors over the entire physical semileptonic region one then
requires reliable and model-independent extrapolation methods constrained by the
available lattice data and any further input quantum-field theory provides. The
ideas presented here have previously been published in Ref.~\cite{Flynn:2023qmi}
and applied in Ref.~\cite{Flynn:2023nhi}.
\section{Fit ansatz}
Boyd, Grinstein and Lebed (BGL) proposed one such model-independent ansatz~\cite{Boyd:1994tt},
\begin{equation}
 \label{eq:BGLparametrisation}
  f_X(q_i^2) = \frac1{B_X(q_i^2)\phi_X(q_i^2,t_0)} \sum_{n=0}^{K_X-1}
  a_{X,n}z(q^2_i)^n= \sum_{n=0}^{K_X-1} Z_{XX,in}a_{X,n},
\end{equation}
with the unitarity constraint for the coefficients, $|{\bf a}_X|^2\le 1$,
derived from dispersion theory. $\phi_X(q^2,t_0)$ is a known ``outer function''
and the Blaschke factor $B_X(q^2)$ is chosen to vanish at the positions of
sub-threshold poles $M^X_{i}$. Similar ideas underlie the approaches in
for example~\cite{Caprini:1997mu, Bourrely:2008za}. The subscript $X$ specifies the
form factor, e.g. $X=+,0$ for the vector and scalar form factor in tree-level
pseudoscalar-to-pseudoscalar decay $B_s\to K\ell\nu$. To the very right of the equation we
introduce a vector-matrix notation for the ansatz. The objective is then to
determine the coefficients $a_{X,n}$ from a finite number $N_{\rm data}$ of
experimental or theory data points.

Within a frequentist fitting strategy the constraint on the number of degrees of
freedom, $N_{\rm dof}=N_{\rm data}-K_X\ge 1$, often limits one's ability to
estimate the truncation error reliably. Moreover, a meaningful implementation
and interpretation of the unitarity constraint within the Frequentist framework
is not straight forward.  Here instead we propose to fit the parameterisation
using Bayesian inference. This provides a conceptually clean way to implement
the unitarity constraint and also to determine coefficients of the
parameterisation beyond the Frequentist bound on $N_{\rm dof}$.  In particular,
we propose to use the unitarity constraint as a regulator for the higher-order
coefficients. In essence, the unitarity constraint forces all coefficients to
lie within a limited $K_X$-dimensional space. Together with the fact that
$|z|\le 1$ in Eq.~(\ref{eq:BGLparametrisation}), this leads to a suppression of
higher-order terms, allowing us to probe from which point on the statistical
error dominates over the truncation error.

\section{An aside on the unitarity constraint}
\begin{figure}
    \begin{center}
    \includegraphics[width=14cm]{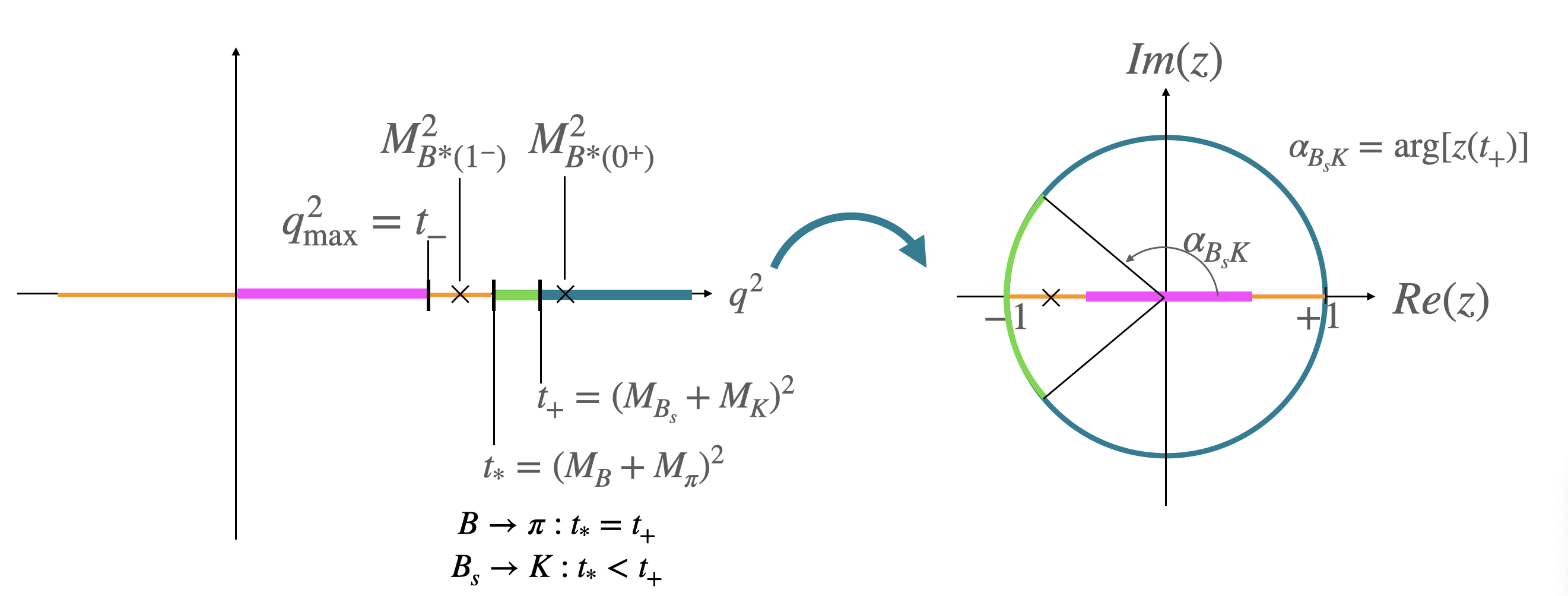}
    \end{center}
    \caption{Illustration of the mapping $q^2 \to z(q^2)$, showing the physical
      semileptonic region (magenta), poles (crosses) and branch cuts (light and
      dark green).}\label{fig:analytic}
\end{figure}
In Fig.~\ref{fig:analytic} we illustrate how the real $q^2$ axis maps onto the
complex unit disk where the $z$ parameter in Eq.~\eqref{eq:BGLparametrisation}
is defined. The physical semileptonic region is mapped to a range on the real
$z$ axis around zero, and the branch cut is mapped onto the unit circle. For the
$b\to u$ transition in $B\to\pi\ell\nu$ and $B_s\to K\ell\nu$ there are two
qualitatively different situations: for $B\to\pi \ell\nu$ the %branch cut
relevant {two-particle production threshold and lower limit of
  integration} for the corresponding unitarity constraint 
  %starts
{is} at $q^2= (M_B+M_\pi)^2$, while the relevant {threshold} for $B_s\to K\ell\nu$  {is}
at $(M_{B_s}+M_K)^2$, which is the part {corresponding to} the
arc $[-\alpha_{B_s K},+\alpha_{B_s K}]$ {in the $z$-plane}, as
indicated in the figure (see
also~\cite{Gubernari:2020eft,Gubernari:2022hxn,Blake:2022vfl}). As a result, the
corresponding unitarity constraints are

\begin{equation}
\frac{1}{2\pi i}\oint_C\frac {dz}{z}\theta_z |B_X(q^2)\phi_X(q^2,t_0)f_X(q^2)|^2\le 1 \,,
\end{equation}
where the step function $\theta_z=\theta(\alpha_{B_s K}-|{\rm arg}[z]|)$ 
in the $B_s\to K$ case restricts
the integration over the unit circle to the relevant segment, \emph{i.e.}\ the one corresponding to the branch cut above the $B_s K$ threshold $t_+$.
Inserting the BGL expansion Eq.~\eqref{eq:BGLparametrisation}, the unitarity constraint takes the compact form~\cite{Flynn:2023qmi}
\begin{equation}
\label{eq:modified unitarity constraint}
\sum\limits_{i,j\ge 0}a_{X,i}^\ast \langle z^i|z^j\rangle_{\alpha_{B_sK}} a_{X,j} \equiv
    |{\bf a}_X|^2_{\alpha_{B_sK}}\le 1\,,
\end{equation}
where the inner product is known analytically,
\begin{equation}
  \langle z^i|z^j\rangle_\alpha =
  \frac 1{2\pi}\int\limits_{-\alpha}^\alpha d\phi (z^i)^\ast
  z^j|_{z=e^{i\phi}}
  = \begin{cases}
    \displaystyle
    \frac{\sin(\alpha(i-j))}{\pi(i-j)} & i\neq j\,,\\
    \displaystyle\frac\alpha\pi & i=j\,.
  \end{cases}
  \label{eq:metric}
\end{equation}

\section{Frequentist fit}
We begin with the discussion of fits to lattice results for the $B_s\to\ K\ell\nu$ scalar $(f_0(q_i^2))$ and vector $(f_+(q_i^2))$ form factors, where we assume that we have data for $N_0$ and $N_+$ data points respectively. We collect all results in the data vector
\begin{equation}
{\bf f}^{\,T}=({\bf f}_+^T,{\bf f}_0^T)=(f_+(q^2_0),f_+(q^2_1),\dots,f_+(q^2_{N_+-1}),
f_0(q^2_0),f_0(q^2_1),\dots,f_0(q^2_{N_0-1}))\,.
\end{equation}
Correspondingly, we collect the BGL parameters into the parameter vector 
\begin{equation}\label{eq:avec definition}
{\bf a}^T=({\bf a}_+^T,{\bf a}_0^T)=(a_{+,0},a_{+,1},a_{+,2},...,a_{+,K_+-1},a_{0,1},...,a_{0,K_0-1})\,.
\end{equation}
With this notation the Frequentist fit is defined as the minimisation of 
\begin{equation}\label{eq:chisq_BGL}
\chi^2({\bf a},{\bf f})=
\left[{\bf f}-Z{\bf a}\right]^T
C_{\bf f}^{-1}
\left[{\bf f}-Z{\bf a}\right]\,,
\end{equation}
with respect to the parameters, where $C_{\bf f}$ is the correlation matrix of the input data ${\bf f}$.
Details on how the kinematical constraint $f_+(0)=f_0(0)$ can be implemented through the matrix
$Z_{XX}$ can be found in Ref.~\cite{Flynn:2023qmi}. Note that we chose to use
this constraint to eliminate the parameter $a_{0,0}$, which is therefore missing
from the definition of ${\bf a}$ in Eq.~\eqref{eq:avec definition}. The solution for the Frequentist fit 
is 
\begin{equation}\label{eq:frequentist_sln_a}
    {\bf a}=\left(Z^T C_{\bf f}^{-1} Z\right)^{-1}ZC_{\bf f}^{-1}{\bf f}\,,\qquad 
      C_{\bf a}=\left(Z^T C_{\bf f}^{-1}Z\right)^{-1}\,,
\end{equation}
where $C_{\bf a}$ is the covariance matrix of the fit parameters ${\bf a}$.
\section{Bayesian fit}
Within the Bayesian framework the fit is defined in terms of the expectation value
\begin{equation}\label{eq:Bayesian main}
\langle g({\bf a})\rangle=\mathcal{N}\int d{\bf a}\,g({\bf a})\,\pi({\bf a}|{\bf f},C_{\bf f})\,\pi_{\bf a}\,,
\end{equation}
where $g({\bf a})$ is a function defined in terms of the BGL parameters. The 
probability density for the integral is given as
\begin{equation}
 \pi_{\bf a}({\bf a}|{\bf f},C_{{\bf f}_p})\pi_{\bf a}\propto{\rm exp}\left(-\frac 12\chi^2({\bf a},{\bf f}_p)\right)\theta(1-|{\bf a}_{+}|_{\alpha}^2)\theta(1-|{\bf a}_{0}|_{\alpha}^2)\,,
\end{equation}
where the two Heaviside functions restrict the integration to parameters compatible with the unitarity constraint. In Ref.~\cite{Flynn:2023qmi} we explain 
how to estimate the integral in~Eq.~\eqref{eq:Bayesian main} by means of sampling from a multivariate normal distribution.

\section{Results for $B_s\to K\ell\nu$}
Table~\ref{tab:Frequ HPQCD14} shows, as an example, the results for Frequentist fits to the HPQCD~14 data of~\cite{Bouchard:2014ypa}. The first two columns indicate the order of the fit in a given row for the vector and scalar form factor, respectively. Acceptable fits (see $p$-values in the 3rd-last column) are achieved only for $K_+\ge 3$ and $K_0\ge 2$. Based on the results in this table, which shows various possible fits with $N_{\rm dof}\ge 1$, solid conclusions about convergence cannot be drawn. 
Tab.~\ref{tab:Bayesian fit HPQCD14} shows the results of the Bayesian fit with $K_{+,0}\le 8$,
where the convergence of the fit parameters is clearly visible. The higher-order coefficients are constrained by the unitarity constraint, which acts as regulator -- without this the parameters would not be well constrained. 
%
%\begin{table}%\normalsize
\begin{table}
\begin{center}
\small
\begin{tabular}{l@{\hspace{1mm}}lllllll@{\hspace{2mm}}l@{\hspace{2mm}}l@{\hspace{2mm}}lllllllllllllllllllllllllllllllllllllllll}
\hline\hline
$K_+$&$K_0$&\multicolumn{1}{c}{$a_{0,0}$}&\multicolumn{1}{c}{$a_{0,1}$}&\multicolumn{1}{c}{$a_{0,2}$}&\multicolumn{1}{c}{$a_{+,0}$}&\multicolumn{1}{c}{$a_{+,1}$}&\multicolumn{1}{c}{$a_{+,2}$}&$p$&$\chi^2_{\rm red}$&$N_{\rm dof}$\\
\hline
2&2&0.0883(44)&-0.250(17)&-         &0.0270(13)&-0.0792(50)&-& 0.03& 2.93&3\\
2&3&0.0880(44)&-0.242(19)&0.053(65) &0.0273(13)&-0.0760(63)&-& 0.02& 4.06&2\\
3&2&0.0906(45)&-0.240(17)&-         &0.0257(14)&-0.0805(50)&0.068(31)& 0.15& 1.89&2\\
3&3&0.0908(46)&-0.215(22)&0.138(71) &0.0262(14)&-0.0727(64)&0.096(34)& 0.97& 0.00&1\\
\hline\hline\\
\end{tabular}\caption{Frequentist fit results for HPQCD 14 data, where $\chi^2_{\rm red}=\chi^2/N_{\rm dof}$.}\label{tab:Frequ HPQCD14}
\end{center}
\end{table}
\begin{table}
\begin{center}
\small
\begin{tabular}{l@{\hspace{1mm}}l@{\hspace{2mm}}l@{\hspace{2mm}}l@{\hspace{2mm}}l@{\hspace{2mm}}l@{\hspace{2mm}}l@{\hspace{2mm}}l@{\hspace{2mm}}l@{\hspace{2mm}}l@{\hspace{2mm}}l@{\hspace{2mm}}llllllllllllllllllllllllllllllllllllllll}
\hline\hline
$K_+$&$K_0$&\multicolumn{1}{c}{$a_{0,0}$}&\multicolumn{1}{c}{$a_{0,1}$}&\multicolumn{1}{c}{$a_{0,2}$}&\multicolumn{1}{c}{$a_{0,3}$}&\multicolumn{1}{c}{$a_{0,4}$}&\multicolumn{1}{c}{$a_{0,5}$}&\multicolumn{1}{c}{$a_{0,6}$}&\multicolumn{1}{c}{$a_{0,7}$}\\%&\multicolumn{1}{c}
%{$a_{0,8}$}\\
%&\multicolumn{1}{c}{$a_{0,9}$}&
\hline
2&2&0.0883(44)&-0.250(17)&- &- &- &- &- &- &\\%- &-&\\
2&3&0.0880(44)&-0.243(19)&0.052(65)&- &- &- &- &- &\\%- &-&\\
3&2&0.0907(46)&-0.240(17)&- &- &- &- &- &- &\\%- &-&\\
3&3&0.0906(44)&-0.215(22)&0.137(73)&- &- &- &- &- &\\%- &-&\\
3&4&0.0907(47)&-0.215(22)&0.14(11)&-0.01(31)&- &- &- &- &\\%- &-&\\
4&3&0.0907(45)&-0.214(22)&0.139(72)&- &- &- &- &- &\\%- &-&\\
4&4&0.0907(46)&-0.215(25)&0.12(19)&-0.08(60)&- &- &- &- &\\%- &-&\\
5&5&0.0909(46)&-0.218(25)&0.10(19)&-0.12(55)&0.04(63)&- &- &- &\\%- &-&\\
6&6&0.0907(45)&-0.217(25)&0.10(19)&-0.11(53)&0.06(66)&-0.02(66)&- &- &\\%- &-&\\
7&7&0.0907(46)&-0.217(26)&0.11(20)&-0.08(51)&0.03(73)&0.03(81)&-0.04(70)&- &\\%- &-&\\
8&8&0.0908(46)&-0.217(25)&0.11(20)&-0.08(50)&-0.01(84)&0.1(1.0)&-0.09(96)&0.08(74)&\\%- &-&\\
%9&9&0.0907(46)&-0.215(25)&0.13(22)&-0.05(50)&-0.06(95)&0.2(1.4)&-0.2(1.5)&0.1(1.2)&-0.05(82)&-&\\
%10&10&0.0907(46)&-0.214(27)&0.15(24)&-0.03(49)&-0.2(1.1)&0.4(1.8)&-0.5(2.2)&0.4(2.1)&-0.3(1.6)&0.13(90)&\\
\hline\hline\\
\end{tabular}
\caption{Results for BGL coefficients for the scalar form factor (results for the vector form factor can be found in~\cite{Flynn:2023qmi}).}\label{tab:Bayesian fit HPQCD14}
\end{center}
\end{table}

While the Frequentist fit can provide important information about the compatibility of the fit function with the data, the Bayesian fit allows for a meaningful imposition of the unitarity constraint and also to study the convergence of the BGL parameterisation. We think that future 
analyses of form-factor data should take advantage of this complementarity. 

\section {$B\to D^\ast \ell\nu$ fit}
The method can be extended to other decay channels such as the semileptonic decay of a pseudo-scalar to a vector particle. We consider the case of $B\to D^\ast \ell\nu$ for which lattice data exists from FNAL/MILC~\cite{FermilabLattice:2021cdg}, HPQCD~\cite{Harrison:2023dzh} and JLQCD~\cite{Aoki:2023qpa}. 
\begin{figure}
    \begin{center}
    \includegraphics[width=16cm]{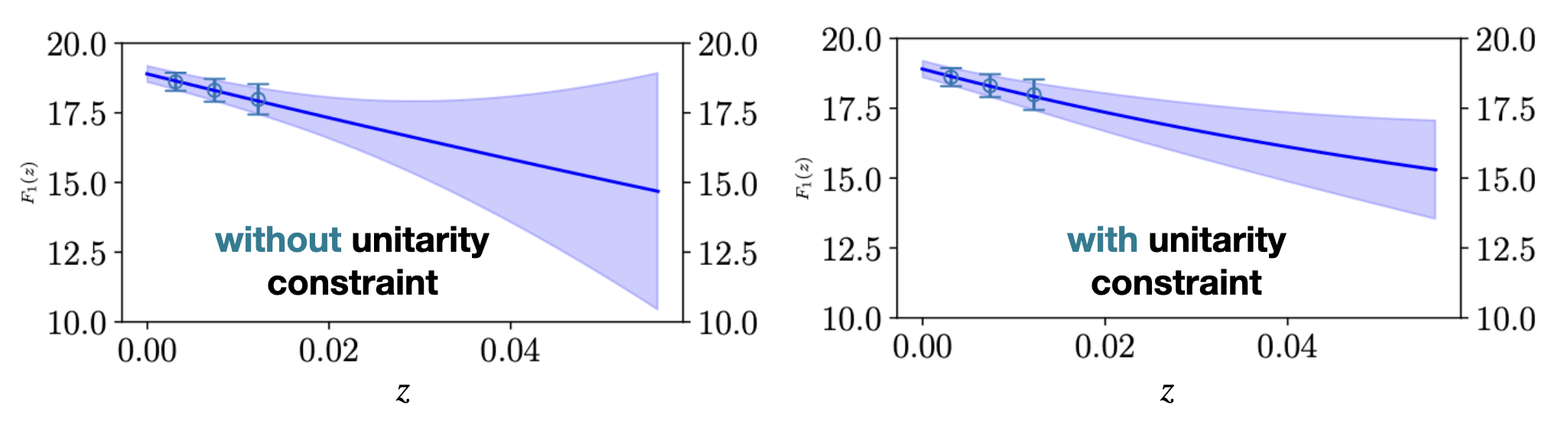}
    \end{center}
    \caption{Result for BGL fit to JLQCD data~\cite{Aoki:2023qpa} for the $B\to D^\ast\ell\nu$
    form factor $\mathcal{F}_1$. Fit without (left) and with (right) unitarity constraint.}
    \label{fig:BtoDstar_F1}
\end{figure}
Fig.~\ref{fig:BtoDstar_F1} shows the result for the form factor $\mathcal{F}_1$ from a simultaneous fit to the JLQCD data over all four form factors
$f,\, g,\,\mathcal{F}_1,\, \mathcal{F}_2$, including kinematical constraints between $\mathcal{F}_1$ and $f$, and $\mathcal{F}_1$ and $\mathcal{F}_2$, respectively. The plots show the fit results once without imposing unitarity and once with. The result of imposing unitarity is a substantial reduction in the statistical error in the extrapolation of the lattice data towards zero momentum transfer (towards the right-hand side of the plots).

\section{Conclusions}
We presented novel ideas for fitting model- and truncation-independent parametersations to data for semileptonic decay form factors. The combination of information from Frequentist and Bayesian fits allows for a comprehensive understanding of fit-quality and truncation dependence. Moreover, the Bayesian framework allows for a meaningful imposition of the unitarity constraint 
to regulate the determination of fit parameters and, as demonstrated for the case of 
$B\to D^\ast \ell\nu$, to reduce the statistical error in form-factor extrapolations. The code underlying the results of~\cite{Flynn:2023qmi} is publicly available under~\cite{fittingPaperCode}.

\bibliographystyle{jhep}
\bibliography{bib}

\end{document}